\PassOptionsToPackage{colorlinks=true
	,urlcolor=blue
	,anchorcolor=blue
	,citecolor=blue
	,filecolor=blue
	,linkcolor=blue
	,menucolor=blue
	,linktocpage=true
	,pdfproducer=medialab
	,pdfa=true}{hyperref}
\documentclass{article}
\pdfoutput=1
\usepackage{jcappub}
\usepackage{graphicx}
\usepackage[latin1]{inputenc}
\usepackage{amsmath,upgreek}
\usepackage{amssymb}
\usepackage{a4wide}
\usepackage{url,hyperref}
\newcommand{\pd}[3]{\frac{\partial^{#3} #1}{\partial {#2}^{#3}}} 
\newcommand{\td}[3]{\frac{d^{#3} #1}{d {#2}^{#3}}} 
\renewcommand{\v}[1]{\ensuremath{\mathbf{#1}}} 
\newcommand{\gv}[1]{\ensuremath{\mbox{\boldmath$ #1 $}}}

\title{Indirect detection prospects for d$^*$(2380) dark matter}
\author{Geoff Beck}

\affiliation{School of Physics, University of the Witwatersrand, Private Bag 3, WITS-2050, Johannesburg, South Africa}

\emailAdd{geoffrey.beck@wits.ac.za}

\abstract{
	A Bose-Einstein condensate of the hexaquark particle known as d$^*$(2380) has been recently proposed as a dark matter candidate by the authors in Bashkanov \& Watts 2020. This particle can produced in an abundant condensate state in the early universe and is argued to satisfy all the stability and weak interaction constraints of a viable dark matter candidate. This dark matter candidate is able to evade direct detection bounds and is suggested to have the best observational prospects in the form of indirect astrophysical emissions due to the decay of the d$^*$ condensate. In this work we test the indirect observational prospects of this form of dark matter and find that its low mass $\sim 2$ GeV mean that sub-GeV gamma-rays searches have the best prospects in the Milky-Way galactic centre where we find $\Gamma_{d^*} < 3.9 \times 10^{-24}$ s$^{-1}$, with current extra-galactic data from M31 and the Coma cluster producing constraints on the d$^*$ decay rate two orders of magnitude weaker. In dwarf galaxies we show that the future GAMMA-400 instrument has the potential to probe down to $\Gamma_{d^*} \sim 10^{-25}$ s$^{-1}$ with 4 years of exposure time.
}

\begin{document}
\maketitle
\section{Introduction}
The nature of Dark Matter (DM) remains one of the most important problems in astroparticle physics. Despite extensive gravitational evidence~\cite{Aghanim:2018eyx,Koopmans_2003,Metcalf_2004,Hoekstra_2002,Moustakas_2003} (or see \cite{Bertone:2018xtm} for a review) neither direct~\cite{xenon2018} (or see \cite{Schumann_2019} and references therein for a review) nor indirect searches~\cite{Bertone:2018xtm,Ackermann_2016,Albert_2017,Colafrancesco2006,Colafrancesco2007,beckm312019,regis2017,Chan_2019,chan2017,HESS:2016ygi,egorov2013,FermiSMC2016,siffert2011,storm2017,gs2016,Beck_2019} have produced anything beyond parameter space constraints on models of cold (or non-relativistic at the time of freeze-out) DM normally favoured by the eponymous, and now standard, $\Lambda$CDM model of cosmology~\cite{Aghanim:2018eyx}. In particular direct detection experiments have strongly narrowed the space of viable supersymmetric Weakly Interacting Massive Particles (WIMPs)~\cite{xenon2018}. Other promising candidates such as sterile neutrinos are being actively searched for in both indirect and laboratory experiments, see \cite{Boyarsky_2019,Adhikari_2017} for reviews. In addition, the axion or axion-like-particle has been the subject of searches in multiple forms~\cite{Duffy_2009,axions2015,cast2017,Braine_2020}. Despite a variety of theoretically well-motivated candidates for a DM particle, and a wide array of on-going searches, no positive evidence has emerged in favour of any particular model. This makes the exploration of new ideas to explain the nature of DM both necessary and worthwhile.  
This work is designed to explore the recent proposal of a new form of light quark based DM. This consists of having the hexaquark particle d$^*$(2380) formed in the early universe and existing until the present epoch in the form of a stable Bose-Einstein Condensate (BEC)~\cite{Bashkanov_2020}. This proposal is particularly interesting as, unlike the aforementioned DM candidates, $d^*$ particles are not part of an extension of the standard model of particle physics. The authors of \cite{Bashkanov_2020} argue that production of $d^*$ particles, during the quark-gluon plasma to hadronic phase transition, would be copious enough to account for present inferred DM abundance~\cite{Aghanim:2018eyx} and that the condensate's interactions with other matter would be sufficiently weak to justify consideration as a DM candidate. 

The authors in \cite{Bashkanov_2020} further argue that the most probable signature of this form of DM is likely to come from astrophysical emissions from the decay of the d$^*$ BEC. In this letter we explore the potential of indirect observations to detect the signatures of d$^*$ decay in both radio and gamma-ray frequencies. This is done by leveraging similar techniques to those used in multi-frequency indirect searches for cold DM particles. The schematic of the idea being that if the consequences of the decay of a single $d^*$ are known, then standard model particle yields from the decay can be used to compute the observable emissions in an astrophysical environment. This requires incorporating data as to the structure of a target DM halo where the decays take place as well as the astrophysical environment in the form of local gas densities and magnetic field strengths, both of which can influence emissions from bremsstrahlung, inverse-Compton scattering, and synchrotron radiation. We specifically explore concentrated DM halos as these environments would contain the largest particle abundances and thus have a larger rate of DM decays, thus making for more visible signatures. 

In particular we will explore several halos as target environments: the Coma cluster of galaxies, the Milky-Way galactic centre (GC), the Reticulum II dwarf galaxy, and M31 or the Andromeda galaxy. All of these environments have been previously used/proposed as targets in indirect DM searches~\cite{egorov2013,beckm312019,gs2016,Colafrancesco2006,Ackermann_2016,Ackermann_2017gc,HESS:2016ygi,Albert_2017,regis2017,bonnivard2015} in a variety of frequency bands from radio to gamma-rays. Barring the Reticulum II dwarf galaxy, all of the chosen targets have been the subject of extensive astrophysical study and thus have relatively well characterised environments in terms of gas densities and magnetic field profiles~\cite{beckm31,briel1992,bonafede2010,ruiz-granados2010}. In addition to this, the structure of each of the DM halos has previously been explored in the literature~\cite{Colafrancesco2006,tamm2012,bonnivard2015,Albert_2017,clumpy_2012,clumpy_2016,clumpy_2018} (and references therein).

For simplicity we assume the decay rate of $d^*$ is a constant $\Gamma_{d^*}$, although it would in principle depend upon the nature of the interactions between cosmic-rays and the $d^*$ BEC~\cite{Bashkanov_2020}.
Our findings are that the best extra-galactic target constraint on the decay rate was from the M31 galaxy with $\Gamma_{d^*} < 10^{-22}$ s$^{-1}$ which corresponds to lower limit on the d$^*$ BEC lifetime several orders of magnitude longer than the age of the universe~\cite{Aghanim:2018eyx}. A model independent search in diffuse galactic gamma-rays from \cite{Essig:2013goa} found $\Gamma < 10^{-24}$ s$^{-1}$ but the photon yield per decay in this model exceeds the hexaquark case from \cite{Bashkanov_2020} by 5 orders of magnitude at energies around $0.1$ GeV where our constraints are derived. So we exceed a `model-translated' version of the \cite{Essig:2013goa} limit by two orders of magnitude in extra-galactic targets. We also note that a very similar model independent limit to that of \cite{Essig:2013goa} is derived from reionisation effects in \cite{Liu_2016}.

In our own galactic centre we find the strongest observational prospect. Using the CLUMPY software~\cite{clumpy_2012,clumpy_2016,clumpy_2018} to determine a dark matter halo density profile we find that $\Gamma_{d^*} < 3.9 \times 10^{-24}$, a full two order of magnitude improvement on extra-galactic data. The galactic centre gamma-ray data set we used was for the Fermi-LAT GeV excess spectrum from \cite{Ackermann_2017gc} as this d$^*$ decay contributes to anomalous emissions not covered by the Fermi templates. The use of this spectrum greatly improves constraints as the flux is at least an order of magnitude below the total signal observed by Fermi-LAT. These limits better the translated limits from \cite{Essig:2013goa} by 5 orders of magnitude and are competitive with the unmodified limits on light DM decay via electrons with final-state radiation (we compare to this case as its spectral shape is most similar to the d$^*$ data from \cite{Bashkanov_2020}). 

This letter is structured as follows: in section~\ref{sec:source} the particle distribution functions from d$^*$ decays are discussed followed by an examination of the emission mechanisms following $d^*$ decay. The DM target halos and their observational data sets are described in \ref{sec:halos}. The results are presented and discussed in section~\ref{sec:results} and conclusions are drawn and summarised in section~\ref{sec:disc}.

\section{The formalism for emissions from d$^*$(2380) decay}
\label{sec:source}

The physical quantity of interest in indirect DM hunts is the flux of photons produced as consequence of the annihilation/decay of DM particles. In the case of the $d^*$ we are interested only in the decay~\cite{Bashkanov_2020}. There are two forms of flux we are interested in and we will label them primary and secondary. Primary emissions result from the prompt production of photons, either directly as part of the decay products or radiation from the prompt decay of these products (and/or final state radiation). Secondary emissions result from the interaction of decay products with the astrophysical environment. Thus, these will consist of Inverse-Compton Scattering (ICS), bremsstrahlung, and synchrotron radiation. All of the secondary radiation studied will be sourced from electrons that result from the products of the $d^*$ decay and thus the processes that affect the energy-loss and diffusion of electrons in astrophysical environments will also be discussed.

\subsection{Primary emissions}
The primary emission flux $S_\gamma$ (photons per unit area per unit time) is simply calculated as follows
\begin{equation}
S_{\gamma} (\nu,r,z) = \int_0^r d^3r^{\prime} \, \frac{Q_{\gamma}(\nu,z,r^\prime)}{4\pi (D_L^2+\left(r^\prime\right)^2)} \; , \label{eq:gamma}
\end{equation} 
where this is the flux from within the volume bounded by radius $r$, $D_L$ is the luminosity distance to the halo centre, and $Q_{\gamma}$ is the gamma-ray source function from $d^*$ decays. 

The source function for particle species $i$ following $d^*$ decay is found according to 
\begin{equation}
Q_i (r,E) = \Gamma_{d^*} \td{N_i}{E}{} \left(\frac{\rho_{d^*}(r)}{m_{d^*}}\right) \; , \label{eq:source}
\end{equation}
where $\Gamma_{d^*}$ is the decay rate, $\td{N_i}{E}{}$ is the particle number produced per unit energy from a d$^*$ decay, $\rho_{d^*} (r)$ is the DM density (which will come from the structure of the DM halo), and $m_{d^*} = 2.380$ GeV. It is worth noting that \cite{Bashkanov_2020} suggest that the condensate decay requires energy injection, via cosmic ray collisions with the $d^*$ BEC for instance, this means that the decay rate is in principle the product of the local cosmic-ray density and the velocity averaged interaction cross-section $\langle \sigma v\rangle$ between the BEC and cosmic rays. In this work we will treat $\Gamma_{d^*}$ as a universal constant for simplicity as the nature of the interactions between $d^*$ and cosmic-rays are not specified in \cite{Bashkanov_2020}. A full treatment will not invalidate the results presented here, rather it will clarify what limits result on $\langle \sigma v \rangle$.
The yield functions $\td{N_i}{E}{}$ are drawn from results presented in \cite{Bashkanov_2020} with the possible decay modes being to photons (via neutral pions) and charged pions. Nucleons and deuterons are also considered in \cite{Bashkanov_2020} but are not relevant in the context of this work. It is important to note that the authors in \cite{Bashkanov_2020} present yields for $\pi^\pm$ but not $e^\pm$, the latter being more relevant for astronomical signatures. Thus, to obtain the $e^\pm$ yields, we convert the pion distributions to those of electrons/positrons following \cite{scanlon1965} which notably omits radiative corrections. An important caveat here is that a pure 6 quark state is assumed for the $d^*$ particle by \cite{Bashkanov_2020} whereas it is argued in the literature that it is likely to be an admixture of hadronic and quark states~\cite{Gal:2019hdb}. This may have implications on the particle source function that cannot be immediately quantified.

\subsection{Electron equilibrium distributions}
\label{sec:electrons}
Secondary emission mechanisms require the spectrum of electrons injected via DM decays. These electrons lose energy through radiation and diffuse from their original point of injection. Thus, their distribution when considering long emission time-scales (as appropriate in astrophysical scenarios) will be taken to be the equilibrium solution to the diffusion-loss equation
\begin{equation}
\begin{aligned}
\pd{}{t}{}\td{n_e}{E}{} = & \; \gv{\nabla} \left( D(E,\v{r})\gv{\nabla}\td{n_e}{E}{}\right) + \pd{}{E}{}\left( b(E,\v{r}) \td{n_e}{E}{}\right) + Q_e(E,\v{r}) \; .
\end{aligned} \label{eq:diff-tot}
\end{equation}
where $Q_e$ is the electron source function from $d^*$ decay and $D$, $b$ are the diffusion and energy-loss functions respectively. The solution method followed here requires the approximation that $D$ and $b$ have no positional dependence, such that
\begin{equation}
D(E) = D_0 \left(\frac{d_0}{1 \; \mbox{kpc}}\right)^{\frac{2}{3}} \left(\frac{\overline{B}}{1 \; \mu\mbox{G}}\right)^{-\frac{1}{3}} \left(\frac{E}{1 \; \mbox{GeV}}\right)^{\frac{1}{3}}  \; , \label{eq:diff}
\end{equation}
where $D_0 = 3.1\times 10^{28}$ cm$^2$ s$^{-1}$, $d_0$ is the magnetic field coherence length, $\overline{B}$ is the average magnetic field strength, and $E$ is the electron energy. The loss-function is found via~\cite{Colafrancesco2006,egorov2013}
\begin{equation}
\begin{aligned}
b(E) & = b_{IC} \left(\frac{E}{1\,\mathrm{GeV}}\right)^2 + b_{sync} \left(\frac{E}{1\,\mathrm{GeV}}\right)^2 \left(\frac{\overline{B}}{1 \; \mu\mbox{G}}\right)^2 \;\\ & + b_{Coul} \left(\frac{\overline{n}}{1 \; \mbox{cm}^{-3}}\right) \left(1 + \frac{1}{75}\log\left(\frac{\gamma}{\left(\frac{\overline{n}}{1 \; \mbox{cm}^{-3}}\right)}\right)\right) + b_{brem} \left(\frac{\overline{n}}{1 \; \mbox{cm}^{-3}}\right) \left(\frac{E}{1\,\mathrm{GeV}}\right)\;,
\end{aligned}
\label{eq:loss}
\end{equation}
where $\overline{n}$ is the average gas density, the coefficients $b_{IC}$, $b_{sync}$, $b_{Coul}$, $b_{brem}$ are the energy-loss rates from ICS, synchrotron emission, Coulomb scattering, and bremsstrahlung. These coefficients are given by $0.25\times 10^{-16}(1+z)^4$/$6.08\times 10^{-16}$ (for CMB/inter-stellar radiation fields), $0.0254\times 10^{-16}$, $6.13\times 10^{-16}$, $4.7\times 10^{-16}$ in units of GeV s$^{-1}$.

The solution to Eq.~(\ref{eq:diff-tot}) when diffusion is negligible is given by~\cite{Colafrancesco2006}
\begin{equation}
\td{n_e}{E}{} = \frac{1}{b(E)} \int_E^{m_\chi} \, dE^{\prime} \, Q_e (r, E^{\prime}) \; .
\end{equation}
When diffusion is not negligible, as in dwarf galaxies like Reticulum II~\cite{Colafrancesco2007,gs2016}, a solution can be found by assuming spherical symmetry~\cite{baltz1999,baltz2004,Colafrancesco2006,Colafrancesco2007}
\begin{equation}
\td{n_e}{E}{} (r,E) = \frac{1}{b(E)}  \int_E^{M_\chi} d E^{\prime} \, G(r,E,E^{\prime}) Q (r,E^{\prime}) \; ,
\end{equation}
where this depends upon the Green's function $G(r,E,E^{\prime})$. This function is expressed as
\begin{equation}
\begin{aligned}
G(r,E,E^{\prime}) = & \frac{1}{\sqrt{4\pi\Delta v}} \sum_{n=-\infty}^{\infty} (-1)^n \int_0^{r_h} d r^{\prime} \; \frac{r^{\prime}}{r_n} \\ & \times \left( \exp\left(-\frac{\left(r^{\prime} - r_n\right)^2}{4\Delta v}\right) - \exp\left(-\frac{\left(r^{\prime} + r_n\right)^2}{4\Delta v}\right) \right)\frac{Q(r^{\prime})}{Q(r)} \; ,
\end{aligned}
\end{equation}
with the sum running over a set of image charges at positions $r_n = (-1)^n r + 2 n r_h$, with $r_h$ being the maximum radius of diffusion under consideration. We follow \cite{Colafrancesco2006} in taking $r_h = 2 R_{vir}$ with $R_{vir}$ being the virial radius of the halo in question. Finally, $\Delta v$ is given by
\begin{equation}
\Delta v =  v(u(E)) - v(u(E^{\prime})) \; ,
\end{equation}
with
\begin{equation}
\begin{aligned}
v(u(E)) = & \int_{u_{min}}^{u(E)} dx \; D(x) \; , \\
u (E) = & \int_E^{E_{max}} \frac{dx}{b(x)} \; . \\ 
\end{aligned}
\end{equation}

\subsection{Secondary emissions}
As the secondary mechanisms are more complicated than primary ones, we will characterise each process $i$ with an emmissivity for frequency $\nu$ and halo position $r$, this being an integral over electron energies $E$:
\begin{equation}
j_{i} (\nu,r,z) = \int_{m_e}^{m_{d^*}} dE \, \left(\td{n_{e^-}}{E}{}(E,r) + \td{n_{e^+}}{E}{}(E,r)\right) P_{i} (\nu,E,r,z) \; ,
\label{eq:emm-he}
\end{equation}
where $\td{n_{e^-}}{E}{}$ is the electron distribution within the source region from d$^*$ decay and $P_i$ is the power emitted at frequency $\nu$ through mechanism $i$ by an electron with energy $E$, at position $r$. The flux produced within a radius $r$ is then found via
\begin{equation}
S_{i} (\nu,r,z) = \int_0^r d^3r^{\prime} \, \frac{j_{i}(\nu,r^{\prime},z)}{4 \pi (D_L^2+\left(r^{\prime}\right)^2)} \; .
\label{eq:flux}
\end{equation}
Thus, in order to characterise each mechanism we need only provide the power $P_i$.

The power produced by the ICS at a photon of frequency $\nu$ from an electron with energy $E$ is given by~\cite{longair1994,rybicki1986}  
\begin{equation}
P_{IC} (\nu,E,z) = c E_{\gamma}(z) \int d\epsilon \; n(\epsilon) \sigma(E,\epsilon,E_{\gamma}(z)) \; ,
\label{eq:ics_power}
\end{equation}
where $E_\gamma (z) = h \nu (1+z)$ with redshift $z$, $\epsilon$ is the energy of the seed photons distributed according to $n(\epsilon)$ (this will taken to be that of the CMB), and
\begin{equation}
\sigma(E,\epsilon,E_{\gamma}) = \frac{3\sigma_T}{4\epsilon\gamma^2}G(q,\Gamma_e) \; ,
\end{equation}
with $\sigma_T$ being the Thompson cross-section and
\begin{equation}
G(q,\Gamma_e) = 2 q \ln{q} + (1+2 q)(1-q) + \frac{(\Gamma_e q)^2(1-q)}{2(1+\Gamma_e q)} \; ,
\end{equation}
with
\begin{equation}
\begin{aligned}
q & = \frac{E_{\gamma}}{\Gamma_e(\gamma m_e c^2 + E_{\gamma})} \; , \\
\Gamma_e & = \frac{4\epsilon\gamma}{m_e c^2} \; ,
\end{aligned}
\end{equation}
where $m_e$ is the electron mass.

The power from bremsstrahlung at photon energy $E_\gamma$ from an electron at energy $E$ is given by~\cite{longair1994,rybicki1986} 
\begin{equation}
P_B (E_{\gamma},E,r) = c E_{\gamma}(z)\sum\limits_{j} n_j(r) \sigma_B (E_{\gamma},E) \; ,
\end{equation}
$n_j$ is the distribution of target nuclei of species $j$, the cross-section is given by
\begin{equation}
\sigma_B (E_{\gamma},E) = \frac{3\alpha \sigma_T}{8\pi E_{\gamma}}\left[ \left(1+\left(1-\frac{E_{\gamma}}{E}\right)^2\right)\phi_1 - \frac{2}{3}\left(1-\frac{E_{\gamma}}{E}\right)\phi_2 \right] \; ,
\end{equation}
with $\phi_1$ and $\phi_2$ being energy dependent factors determined by the species $j$(see \cite{longair1994,rybicki1986} ).

The power from synchrotron emission at frequency $\nu$ from an electron at energy $E$ is given by~\cite{rybicki1986,longair1994}
\begin{equation}
P_{synch} (\nu,E,r,z) = \int_0^\pi d\theta \, \frac{\sin{\theta}}{2}2\pi \sqrt{3} r_e m_e c \nu_g(r) F_{synch}\left(\frac{\kappa(r)}{\sin{\theta}}\right) \; ,
\label{eq:power}
\end{equation}
where $m_e$ is the electron mass, $\nu_g (r) = \frac{e B(r)}{2\pi m_e c}$ is the non-relativistic gyro-frequency, $B(r)$ is magnetic field strength at $r$, $r_e = \frac{e^2}{m_e c^2}$ is the classical electron radius, and the quantities $\kappa(r)$ and $F_{synch}$ are defined as
\begin{equation}
\kappa(r) = \frac{2\nu (1+z)}{3\nu_g(r) \gamma^2}\left[1 +\left(\frac{\gamma \nu_p(r)}{\nu (1+z)}\right)^2\right]^{\frac{3}{2}} \; ,
\end{equation}
with the plasma frequency $\nu_p(r) \propto \sqrt{n_e(r)}$ with $n_e(r)$ being the gas density at $r$, $\gamma$ as the electron Lorentz factor, and
\begin{equation}
F_{synch}(x) = x \int_x^{\infty} dy \, K_{5/3}(y) \approx 1.25 x^{\frac{1}{3}} \mbox{e}^{-x} \left(648 + x^2\right)^{\frac{1}{12}} \; .
\end{equation}

\section{Halo environments and their observed fluxes}
\label{sec:halos}
With the emission mechanisms now detailed we still require some information to compute the $d^*$ decay flux: namely a $d^*$ density $\rho_{d^*}$ as well as magnetic field and gas profiles, $B(r)$ and $n_e(r)$ respectively, for the decay environment. We choose to study dense DM halos, as if $d^*$ constitutes DM, these will present the strongest signatures due to the scaling of Eq.~(\ref{eq:source}) with $\rho_{d^*}$. Thus, in this section, we will enumerate the environmental properties of each halo of interest in this work. Furthermore, we will note what observational data will be compared to predicted $d^*$ emissions in order to place limits on the decay rate $\Gamma_{d^*}$. 

The primary halo characteristic is the DM density profile, entering into flux calculations via Eq.~(\ref{eq:source}). We will make use of the Navarro-Frenk-White (NFW) case~\cite{nfw1996}, the cored Burkert profile~\cite{burkert1995}, and the Einasto profile~\cite{einasto1968}
\begin{equation}
\begin{aligned}
\rho_{nfw}(r)=\frac{\rho_{s}}{\frac{r}{r_s}\left(1+\frac{r}{r_s}\right)^{2}} \; , \\
\rho_{burk}(r)=\frac{\rho_s}{\left(1+ \frac{r}{r_s}\right)\left(1+\left[\frac{r}{r_s}\right]^2\right)} \; ,\\
\rho_{ein}(r)=\rho_{s} \exp\left[-\frac{2}{\alpha} \left(\left[\frac{r}{r_s}\right]^{\alpha} - 1\right)\right] \; ,
\end{aligned}
\label{eq:density}
\end{equation}
where $\alpha$ is the Einasto parameter, while $r_s$ and $\rho_s$ are the characteristic halo scale and density respectively. All of these profiles are used under the assumption that the halo is spherically symmetric.

\subsection{Coma}
In the Coma galaxy cluster we follow \cite{coma-halo-2003} in using the halo parameters $M_{vir} = 1.24 \times 10^{15}$ M$_{\odot}$, $R_{vir} = 3.0$ Mpc, and $r_s = 0.333$ Mpc with an NFW profile ($\rho_s$ is chosen to normalise the density to $M_{vir}$ within $R_{vir}$).

The gas density within the halo is taken to have a profile 
\begin{equation}
n_e(r) = n_0 \left(1 + \left[\frac{r}{r_d}\right]^2\right)^{-q_e} \; ,
\end{equation}
with $n_0 = 3.49 \times 10^{-3}$ cm$^{-3}$, $q_e = 0.981$, $r_d = 0.256$ Mpc from \cite{chen-clusters-2007} corrected to $H_0 = 67.74$.

The magnetic field model is then
\begin{equation}
B(r) = B_0 \left(\frac{n_e(r)}{n_0}\right)^{q_b} \; ,
\end{equation}
with $B_0 = 4.7$ $\mu$G, and $q_b = 0.5$ from \cite{bonafede2010}.

In the Coma galaxy cluster we will compare predicted $d^*$ decay emissions to the diffuse radio data set from \cite{coma-radio2003} and the gamma-ray limits from \cite{Ackermann_2016}. The predicted DM flux $S(\nu)$ will be integrated out to a radius of 30 arcminutes to match the radio contours used in \cite{coma-radio2003}. 

\subsection{M31}
In M31 we will use an NFW profile with parameters found in \cite{tamm2012} to be $M_{vir} = 1.04 \times 10^{12}$ M$_{\odot}$, $R_{vir} = 0.207$ Mpc, and $r_s = 0.0167$ Mpc  ($\rho_s$ is chosen to normalise the density to $M_{vir}$ within $R_{vir}$).
In M31 we use an exponential gas density
\begin{equation}
n_e (r) = n_0 \exp\left(-\frac{r}{r_d}\right) \; ,
\end{equation}
with $n_0 = 0.06$ cm$^{-3}$~\cite{beckm31}, and $r_d \approx 5$ kpc fitted in \cite{ruiz-granados2010}.
In this environment we use the magnetic field model from \cite{ruiz-granados2010} which has, within $r \leq 40$ kpc,
\begin{equation}
B(r) = \frac{4.6 \left(\frac{r_1}{1\,\mathrm{kpc}}\right) + 64}{\left(\frac{r_1}{1\,\mathrm{kpc}}\right) + \left(\frac{r}{1\,\mathrm{kpc}}\right)} \, \mu\mathrm{G} \; ,
\end{equation}
with $r_1 = 200$ kpc.

For testing $d^*$ decay predictions in M31 we make use of the radio frequency data set from \cite{beckm312019}, these are divided into $50'$ and $15'$ observing regions (see \cite{beckm312019} and references therein for further details). For gamma-rays in M31 we use the data points (but not upper-limits at higher energies) from \cite{Ackermann_2017}. We match the integration radius for the DM flux $S(\nu)$ to the region of interest for the data sets we are comparing to.

\subsection{Reticulum II}
The Reticulum II dwarf galaxy is taken to have a Burkert density profile following arguments from \cite{walker2009,adams2014}. This profile has $r_s = 0.139$ kpc~\cite{regis2017} and is normalised to the annihilation J-factor $8 \times 10^{18}$ GeV$^2$ cm$^{-5}$ found in \cite{Albert_2017} within $0.5^\circ$ of the galaxy centre. 

In Reticulum II we follow \cite{regis2017} in using the profiles for gas density and magnetic field strength:
\begin{equation}
n_e (r) = n_0 \exp\left(-\frac{r}{r_d}\right) \; ,
\end{equation}
and
\begin{equation}
B (r) = B_0 \exp\left(-\frac{r}{r_d}\right) \; .
\end{equation}
We take $r_d$ to be given by the stellar-half-light radius with a value of $15$ pc~\cite{bechtol2015,koposov2015} and we assume $B_0 \approx 1$ $\mu$G, $n_0 \approx 10^{-6}$ cm$^{-3}$.

In Reticulum II we use, as our observational data, the diffuse flux limit of $12$ $\mu$Jy at $1.873$ GHz from \cite{regis2017} and the gamma-ray upper-limits from \cite{Albert_2017}. 

\subsection{Galactic centre}
We use only the gamma-ray spectrum in the galactic centre and so our model data consist of just the annihilation J-factor. This J-factor is defined 
\begin{equation}
J(\Delta\Omega) = \int_{\Delta \Omega}\int_{\mathrm{l.o.s}} \rho^2_{d^*}(r(l)) d l d\Omega \; ,
\end{equation}
where $\Delta \Omega$ is the observed solid angle and l.o.s signifies the line of sight. It is evident that this quantity is equivalent to the integral over $r$ in Eq.~(\ref{eq:gamma}).
We find the J-factor, using CLUMPY~\cite{clumpy_2012,clumpy_2016,clumpy_2018}, to be $9.94\times 10^{22}$ GeV$^2$ cm$^{-5}$ within 10$^\circ$ of the galactic centre, assuming an Einasto halo profile. We normalise our halo profile to the stated J-factor (using the CLUMPY halo parameters) in order to determine the decay products from this target. We stress that we do not use the J factor to calculate the decay-based flux, only to normalise the halo profile.

In this environment we test the gamma-ray spectrum of the Fermi-LAT excess from \cite{Ackermann_2017gc} against DM predictions using the $10^\circ$ region of interest around the galactic centre.

\section{Results and discussion}
\label{sec:results}
The results will be presented in the form of spectra from $d^*$ decay for each target halo. This will be in the form of the flux $E(\nu) S(\nu)$ (calculated as the sum of  Eq.~(\ref{eq:gamma}) and each mechanism from Eq.~(\ref{eq:flux})) which has units of energy per unit area per unit time with $E(\nu) = h \nu$. This will be compared with observational data listed for each halo in section~(\ref{sec:halos}). 
We will display predicted spectra assuming a fiducial value of the $d^*$ decay rate $\Gamma_{d^*} = 10^{-24}$ s$^{-1}$. Limits on the actual value of $\Gamma_{d^*}$ are then derived by requiring that the predicted $d^*$ flux is smaller than the observed flux (as $d^*$ cannot contribute more flux than has already been observed, it could only be responsible for some fraction of it). Note that $\Gamma_{d^*}$ being a constant, it acts as a normalisation factor multiplying the calculated spectrum, as displayed in Eq.~(\ref{eq:source}). So the flux for a given $\Gamma_{d^*}$ is found by multiplying the displayed fiducial fluxes $E S(\nu)$ by $\frac{\Gamma_{d^*}}{10^{-24} \, \mathrm{s}^{-1}}$.

The fiducial decay rate is chosen to be similar to the most comparable results in the literature: largely model independent sets from \cite{Essig:2013goa,Liu_2016} which constrain decaying DM models via direct decay to either photons or electrons and positrons. In our case the most comparable source functions from \cite{Essig:2013goa,Liu_2016} are those for DM decaying to electrons and positrons, as the spectral shape resembles those found in this work. However, the limits from \cite{Essig:2013goa,Liu_2016} are not completely compatible with the model from \cite{Bashkanov_2020}. This is because, while the spectral shapes are somewhat similar, the model-independent gamma-ray yields are around 5 orders of magnitude larger than those presented for the decay of d$^*$ particles. This is due to the fact that $d^*$ decays to $e^\pm$ and photons via intermediate states. So a rough benchmark to compare with our limits derived here will be $\Gamma_{d^*} < 10^{-19}$ s$^{-1}$. Additionally, this makes a reasonable benchmark due to the similarity of this $1/\Gamma_{d^*}$ to the age of the universe~\cite{Aghanim:2018eyx}.   

Each spectrum has 3 characteristic peaks: one for each of the synchrotron, ICS, and primary emissions (in ascending frequency order). The synchrotron peak falls below 10 MHz and so will not be displayed (as it is likely not visible with Earth-based radio telescopes). The bremsstrahlung contributes the bridging region between ICS and primary peaks. In each case the synchrotron emission dominates spectrum below a few hundred MHz, with the ICS region from $10^3$ to $10^{11}$ MHz, bremsstrahlung between $10^{11}$ and $10^{15}$ MHz, and finally primary emissions at higher frequencies. These regions vary slightly between each halo as they depend upon the gas and magnetic field strength profiles through the expressions for mechanism powers in section~\ref{sec:source} and also via dependence on electron equilibrium distributions.

\begin{figure}[ht!]
	\centering
	\resizebox{0.8\hsize}{!}{\includegraphics{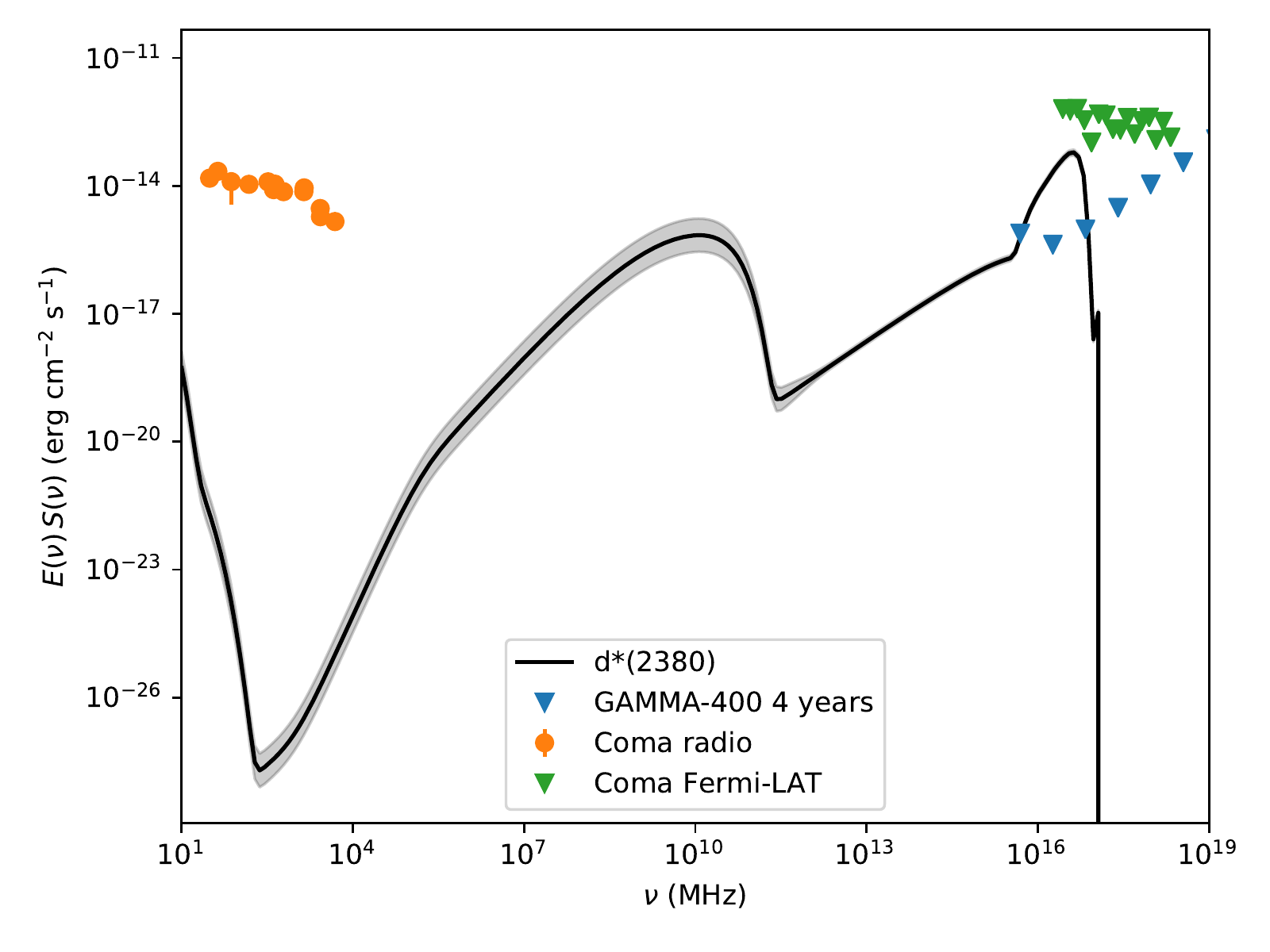}}
	\caption{Predicted multi-frequency spectrum from d$^*$ decay in the Coma galaxy cluster with $\Gamma_{d^*} = 10^{-24}$ s$^{-1}$. The shaded region indicates uncertainties from the magnetic field and halo mass.}
	\label{fig:coma}
\end{figure}

In the case of the Coma galaxy cluster in Figure~\ref{fig:coma} it is evident that existing gamma-ray upper-limits provide the strongest option for constraint in this environment with $\Gamma_{d^*} \leq 2.0 \times 10^{-23}$ s$^{-1}$ to avoid exceeding the limits at $2\sigma$ confidence interval. This is because of the proximity between these data points and the spectrum plotted with the fiducial decay rate. The observed radio data points are at least 5 orders of magnitude above the predicted spectrum and would thus only be exceeded by the predicted spectrum with a decay rate at least $10^5$ times higher than the fiducial value. The resulting constraints from this data are $\Gamma_{d^*} \leq 9.6 \times 10^{-17}$ s$^{-1}$. Our results in the Coma case already exceed the model-translated limits from \cite{Essig:2013goa,Liu_2016} by around 2 orders of magnitude. To supplement the higher energy measurements we have also considered the sensitivity projections of the GAMMA-400 telescope~\cite{g4002013,g4002019} for 4 years of exposure time. This curve indicates that the Coma cluster can be probed for $d_*$ decays to a level of $\Gamma_{d^*} \sim 10^{-26}$ s$^{-1}$ at the 95\% confidence interval, though contamination by other emissions will likely limit this potential.

\begin{figure}[ht!]
	\centering
	\resizebox{0.8\hsize}{!}{\includegraphics{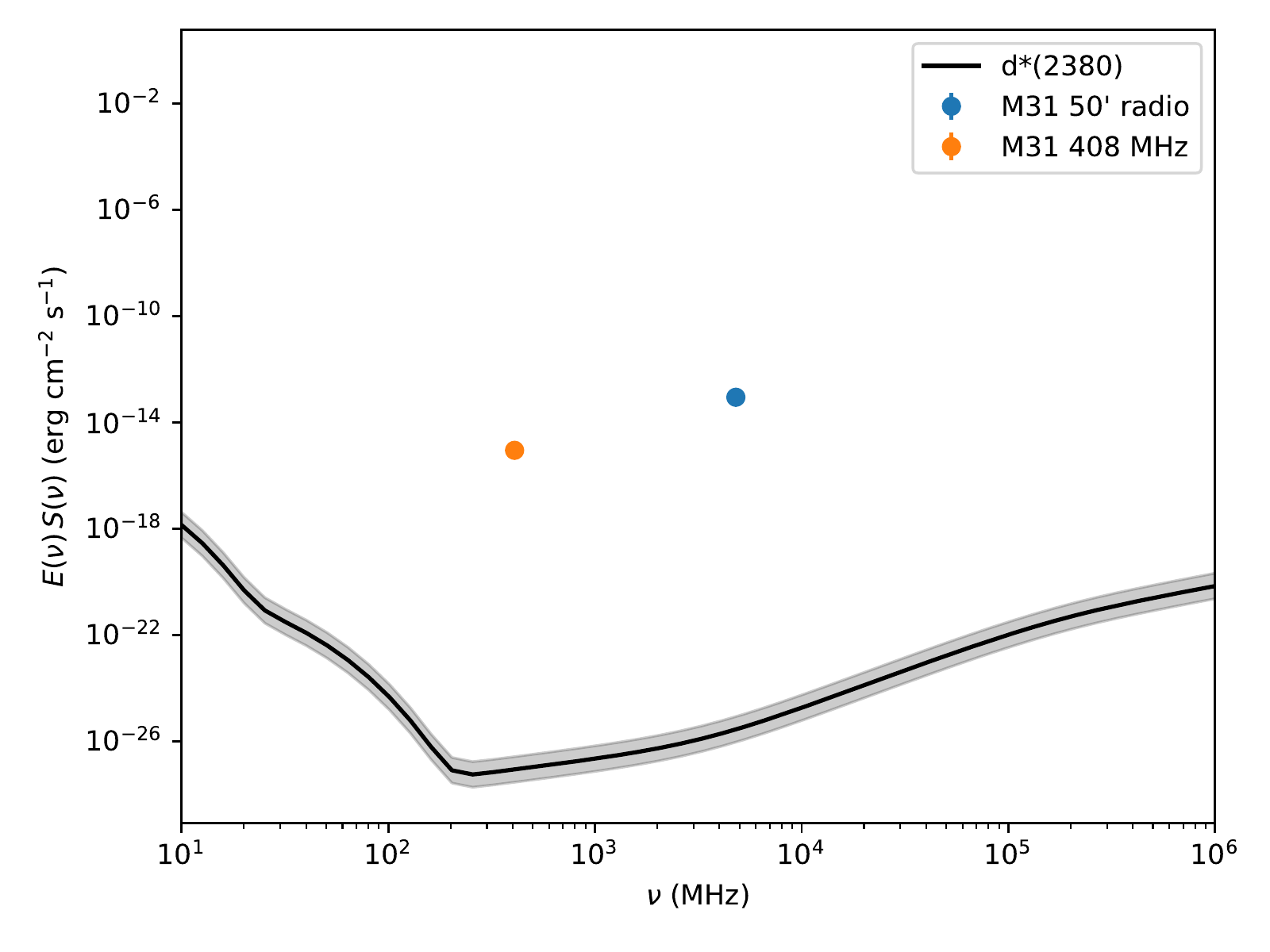}}
	\caption{Predicted radio spectrum from d$^*$ decay in the M31 galaxy within $50'$ with $\Gamma_{d^*} = 10^{-24}$ s$^{-1}$. The shaded region indicates uncertainties from the magnetic field and halo mass.}
	\label{fig:m31-radio}
\end{figure}

\begin{figure}[ht!]
	\centering
	\resizebox{0.8\hsize}{!}{\includegraphics{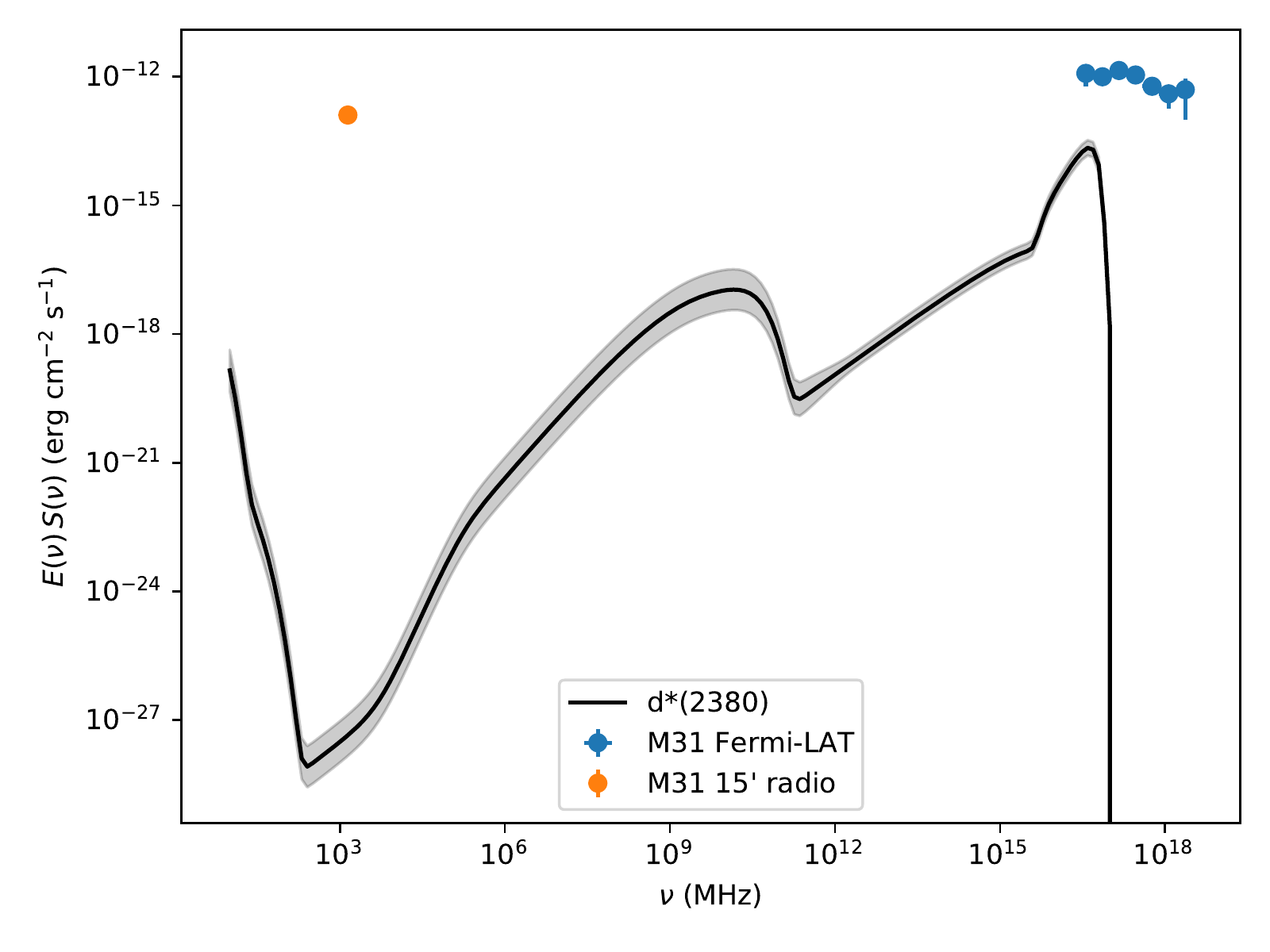}}
	\caption{Predicted multi-frequency spectrum from d$^*$ decay in the M31 galaxy within $15'$ with $\Gamma_{d^*} = 10^{-24}$ s$^{-1}$. The shaded region indicates uncertainties from the magnetic field and halo mass.}
	\label{fig:m31-multi}
\end{figure}

M31 results are displayed in Figs.~\ref{fig:m31-radio} and \ref{fig:m31-multi} are also well below the observed data points for the fiducial $\Gamma$ value. However, the gamma-ray data points from \cite{Ackermann_2017} provide a constraint that $\Gamma_{d^*} \leq 1.2 \times 10^{-22}$ s$^{-1}$ at the 2$\sigma$ confidence level. This is vastly better than the Coma cluster radio constraint, and a factor of 10 or so weaker than the gamma-ray case. 
Substantially weaker limits result from the radio data points, as opposed to higher energies, largely as the synchrotron emissions peak at such a low frequency as a consequence of the $\sim 2$ GeV d$^*$ mass. 

\begin{figure}[ht!]
	\centering
	\resizebox{0.8\hsize}{!}{\includegraphics{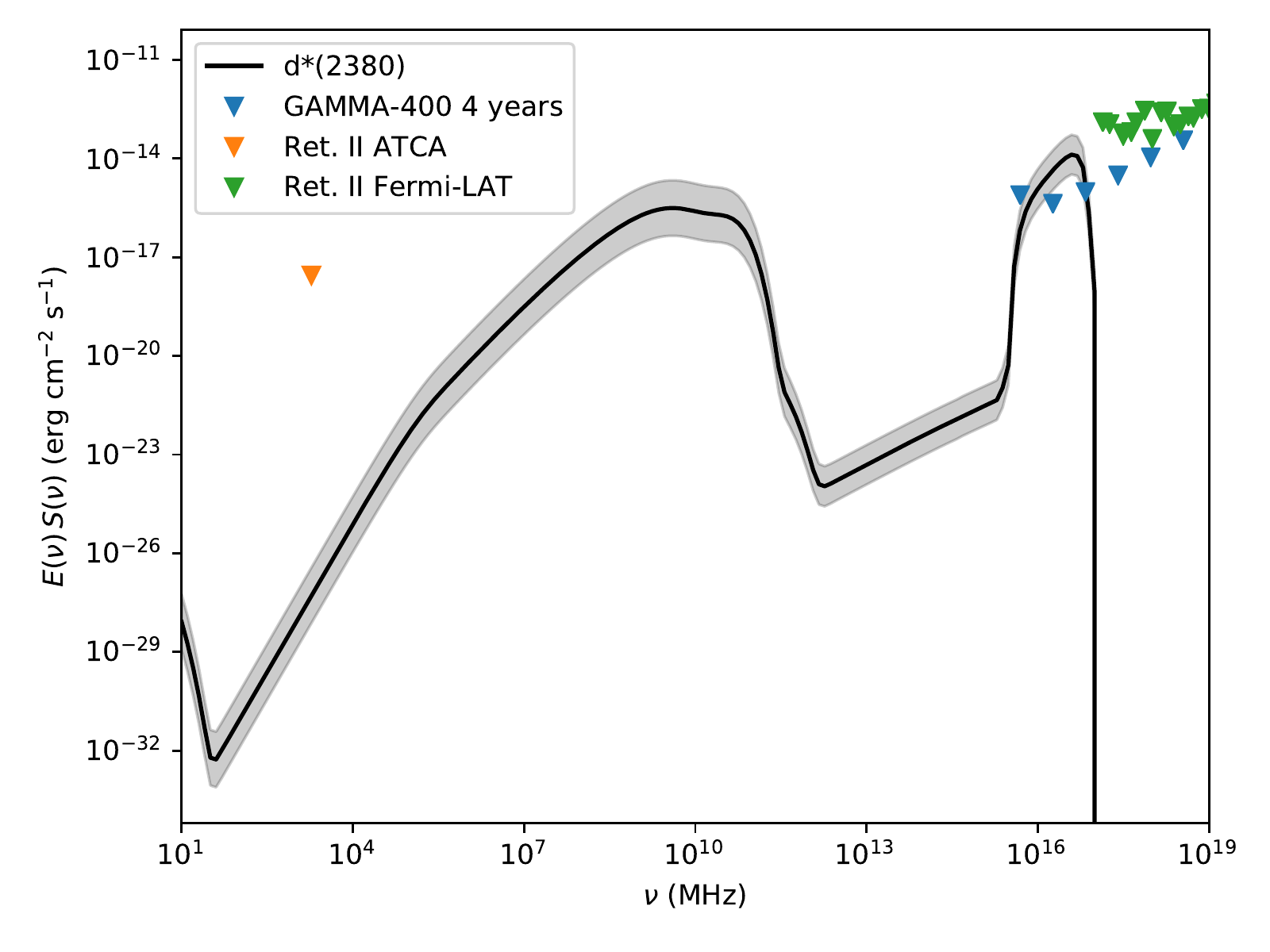}}
	\caption{Predicted multi-frequency spectrum from d$^*$ decay in the Reticulum II dwarf galaxy within $30'$ with $\Gamma_{d^*} = 10^{-24}$ s$^{-1}$. The shaded region indicates uncertainties from the magnetic field and halo mass.}
	\label{fig:ret2}
\end{figure}

In Figure~\ref{fig:ret2} we display the Reticulum II predicted spectrum. Importantly, the gamma-ray data points do not overlap with the predicted spectrum so can provide no constraints. However, the steep gamma-ray peak suggests that lower energy observations could provide a relatively strong probe of d$^*$ decays in this target, as evidence we see the GAMMA-400 sensitivity curve which can result in a limit of $\Gamma_{d^*} < 10^{-25}$ s${-1}$ at the 95\% confidence interval. If we consider just a case where the power-law trend of the Fermi-LAT limits is continued, a $\Gamma_{d^*} \leq 10^{-23}$ s$^{-1}$ is potentially attainable, on par with the Coma cluster. The relatively low energy threshold for ICS dominance in the Reticulum II spectrum also produces surprisingly strong radio limits (given the weak magnetic field assumptions) with $\Gamma_{d^*} \leq 1.04 \times 10^{-14}$ s$^{-1}$.  

\begin{figure}[ht!]
	\centering
	\resizebox{0.8\hsize}{!}{\includegraphics{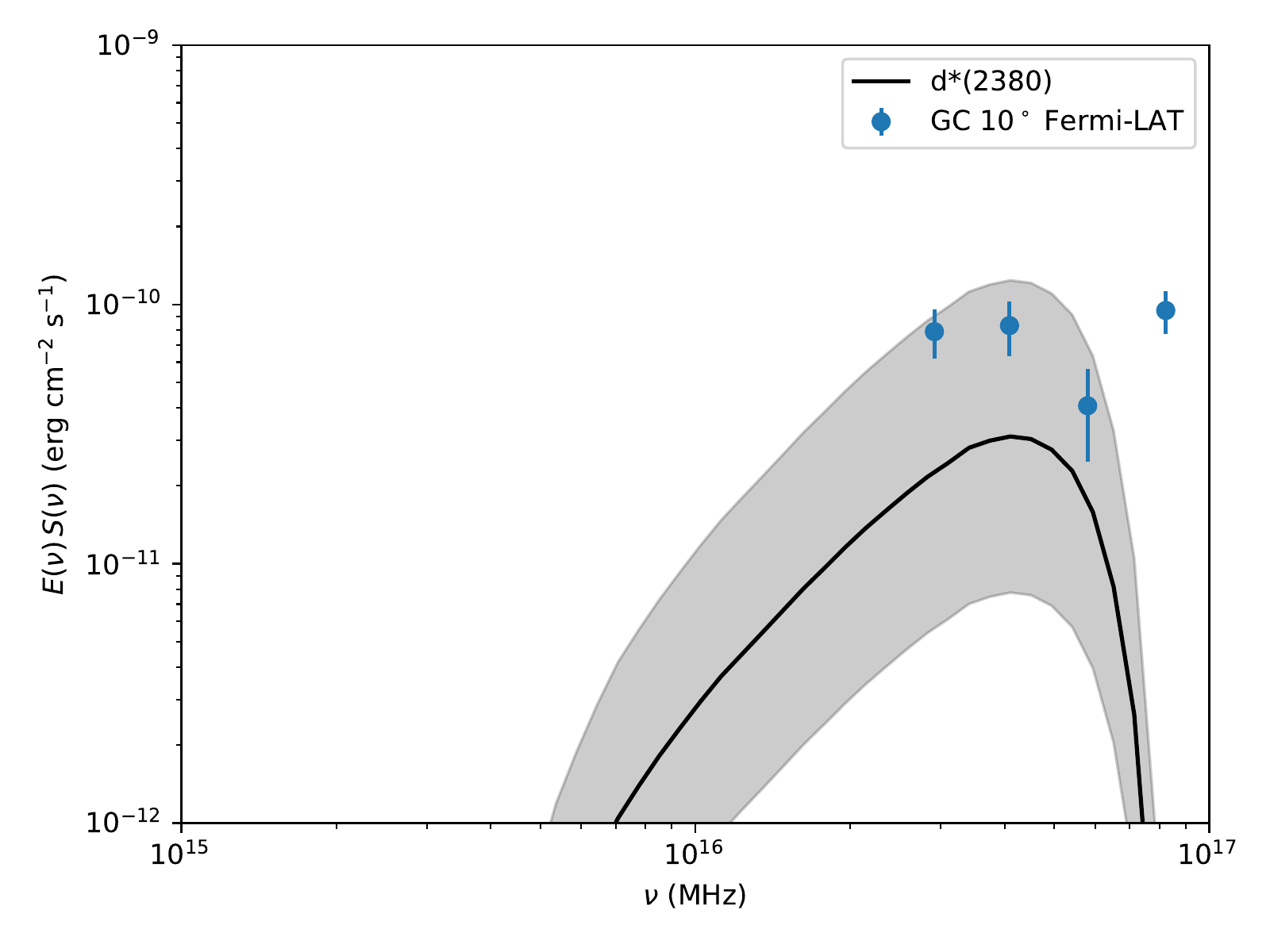}}
	\caption{Predicted multi-frequency spectrum from d$^*$ decay within $10^\circ$ of the Milky-Way galactic centre with $\Gamma_{d^*} = 10^{-24}$ s$^{-1}$. The shaded region indicates uncertainties from the halo J-factor.}
	\label{fig:gc}
\end{figure}

In figure~\ref{fig:gc} we display the case of the galactic centre gamma-ray spectrum compared to the data from the Fermi-LAT diffuse gamma-ray excess within 10$^\circ$ from \cite{Ackermann_2017gc}. Here we see that the $\Gamma_{d^*} = \times 10^{-24}$ s$^{-1}$ case only lies slightly below the observed spectrum and a resulting constraint is that $\Gamma_{d^*} \leq 3.9 \times 10^{-24}$ s$^{-1}$, improving on the model translated value from \cite{Essig:2013goa,Liu_2016} by around 5 orders of magnitude making this competitive even with the unmodified model independent limits. 

\begin{table}[ht!]
	\centering
	\begin{tabular}{|l|l|}
		\hline
		Data set & Limit \\
		\hline
		GC gamma-ray & $\Gamma_{d^*} \leq 3.9 \times 10^{-24}$ s$^{-1}$\\
		M31 gamma-ray & $\Gamma_{d^*} \leq 1.2 \times 10^{-22}$ s$^{-1}$\\
		Coma gamma-ray & $\Gamma_{d^*} \leq 2.0 \times 10^{-23}$ s$^{-1}$\\
		Coma radio & $\Gamma_{d^*} \leq 9.6 \times 10^{-17}$ s$^{-1}$\\
		Reticulum II radio & $\Gamma_{d^*} \leq 1.04 \times 10^{-14}$ s$^{-1}$\\
		M31 radio & $\Gamma_{d^*} \leq 4.9 \times 10^{-14}$ s$^{-1}$\\
		\hline
	\end{tabular}
\caption{Summary of $2\sigma$ confidence interval limits on the d$^*$ decay rate in different DM halos.}
\label{tab:summary}
\end{table}
We present a summary of the constraints on the d$^*$ decay rate $\Gamma_{d^*}$ in Table~\ref{tab:summary}. These results indicate that the various fluxes from d$^*$ decay are weak but benchmarking against the viability of the hexaquark DM model is difficult without a result for $\Gamma$ which would be necessary to reproduce the present-day DM abundance. The most optimistic observation case is from our results is the Milky-Way galactic centre which produces limits from gamma-ray fluxes similar to model-independent limits from \cite{Essig:2013goa,Essig:2013goa} and 5 orders of magnitude better than the roughly translated model-dependent value $10^{-19}$ s$^{-1}$. Despite being two orders of magnitude smaller than $\Gamma_{d^*}$ limits from the galactic centre, M31 and Coma still provide limits $\Gamma_{d^*} \lesssim 10^{-22}$ s$^{-1}$ and $\Gamma_{d^*} \lesssim 10^{-23}$ s$^{-1}$ respectively, indicating a particle lifetime lower limit around two orders of magnitude in excess of the age of the universe. If not for the limited energy range of the Reticulum II gamma-ray data it is likely that the strongest extra-galactic environment for these constraints is dwarf galaxies (particularly with gamma-ray measurements from instruments like GAMMA-400~\cite{g4002013,g4002019,g4002020} or perhaps millimetre telescopes). Radio limits are weak across the board due to the very low energy of the synchrotron peak as a consequence of the small d$^*$ mass. Particularly, this synchrotron peak lies below 10 MHz and is therefore likely unobservable within the atmosphere of Earth. X-ray observations are not considered here as they tend to fall into the trough between the ICS and gamma-ray emission peaks (as can be seen for THESEUS~\cite{Amati_2018} for example) and thus have great difficulty in probing these models.

\section{Summary and conclusions}
\label{sec:disc}

We have presented a comparison of emission predictions from the decay of $d^*$ hexaquark particles to observed multi-frequency spectra in M31, the Milky-Way galactic centre, Reticulum II, and the Coma galaxy cluster. These results were used to place the first limits on the decay rate of the $d^*$ particle and explore the suggestion in \cite{Bashkanov_2020} that indirect astrophysical observations would provide the strongest signatures of $d^*$ BEC DM. Our findings were that the best limit on the decay rate is found in the Milky-Way galactic centre with $\Gamma_{d^*} \leq 3.9 \times 10^{-24}$ s$^{-1}$ by comparing predicted spectra to the Fermi-LAT excess spectrum from \cite{Ackermann_2017gc}. 

Since this DM model has not been probed indirectly before it is not trivial to compare to existing literature. This is because any relevant limits will have to be sourced from model independent studies. The results most comparable to our work yield $\Gamma \leq \times 10^{-24}$ s$^{-1}$~\cite{Essig:2013goa,Liu_2016} for light DM (with similar mass to $d^*$) decaying into $e^\pm$ directly. This approach results in a similar spectral shape to ours but over-produces the $d^*$ photon spectrum by a factor of $10^5$. Thus, our results are comparable to existing model-independent limits at face value but also greatly exceed a `model adjusted' version of the \cite{Essig:2013goa} limit at $\Gamma \leq \times 10^{-19}$ s$^{-1}$. The discrepancy between the model independent case and that of $d^*$ arises because $d^*$ decays produce stable particles via initial decay to pions, rather than the direct production explored in the model independent literature in question.

The uncertainties from halo parameters in the presented results are notably small in comparison to the difference with the lower limits from \cite{Essig:2013goa} (model independent limits on decaying DM) and tend to be smaller in the simpler gamma-ray region of the spectrum. However, unquantified uncertainties exist in terms of the required $\Gamma$ value to achieve a significant present-day DM fraction and in the formalism used to convert charged pion products to electrons/positrons (as the quantitative effect of the neglected radiative corrections is unknown). This latter uncertainty does not affect the gamma-ray results as they were all attained from the neutral pion decay channels of d$^*$.

Despite these uncertainties it seems likely that the emissions from d$^*$ hexaquark decay produce relatively weak fluxes in astrophysical environments (a similar conclusion is reached in \cite{Chan_2020}) but that gamma-ray searches in galaxy clusters, galaxies, and dwarf galaxies at energies below 1 GeV may be able to provide further constraints. In particular, the GAMMA-400 instrument is highlighted as being promising in this regard being able to exclude decay-rates $\Gamma_{d^*} \gtrsim 10^{-25}$ s$^{-1}$ by non-observation with 4 years of exposure time. The data used in this work was not optimised for this kind of indirect DM search so the results produced could undoubtedly be improved upon in the future. One possible improvement in methodology would be to disentangle the constituents of decay rate (as described in section~\ref{sec:source}) in order to produce limits on the $d^*$-cosmic-ray interaction cross-section.

\section*{Acknowledgements}
G.B acknowledges support from a National Research Foundation of South Africa Thuthuka grant no. 117969. This research has made use of the NASA/IPAC Extragalactic Database (NED), which is operated by the Jet Propulsion Laboratory, California Institute of Technology, under contract with the National Aeronautics and Space Administration. This work also made use of the WebPlotDigitizer\footnote{\url{http://automeris.io/WebPlotDigitizer/}}.

\bibliographystyle{JHEP}
\bibliography{dstar}

\providecommand{\href}[2]{#2}\begingroup\raggedright\begin{thebibliography}{10}

\bibitem{Aghanim:2018eyx}
{\scshape Planck Collaboration} collaboration, \emph{{Planck 2018 results. VI.
  Cosmological parameters}},
  \href{https://arxiv.org/abs/1807.06209}{{\ttfamily 1807.06209}}.

\bibitem{Koopmans_2003}
L.~V.~E. Koopmans and T.~Treu, \emph{The structure and dynamics of luminous and
  dark matter in the early-type lens galaxy of 0047-281 at z = 0.485},
  \href{https://doi.org/10.1086/345423}{\emph{The Astrophysical Journal}
  {\bfseries 583} (2003) 606}.

\bibitem{Metcalf_2004}
R.~B. Metcalf, L.~A. Moustakas, A.~J. Bunker and I.~R. Parry,
  \emph{Spectroscopic gravitational lensing and limits on the dark matter
  substructure in q2237+0305}, \href{https://doi.org/10.1086/383243}{\emph{The
  Astrophysical Journal} {\bfseries 607} (2004) 43}.

\bibitem{Hoekstra_2002}
H.~Hoekstra, H.~Yee and M.~D. Gladders, \emph{Current status of weak
  gravitational lensing},
  \href{https://doi.org/10.1016/s1387-6473(02)00245-2}{\emph{New Astronomy
  Reviews} {\bfseries 46} (2002) 767}.

\bibitem{Moustakas_2003}
L.~A. Moustakas and R.~B. Metcalf, \emph{Detecting dark matter substructure
  spectroscopically in strong gravitational lenses},
  \href{https://doi.org/10.1046/j.1365-8711.2003.06055.x}{\emph{Monthly Notices
  of the Royal Astronomical Society} {\bfseries 339} (2003) 607}.

\bibitem{Bertone:2018xtm}
G.~Bertone and M.~P. Tait, Tim, \emph{{A new era in the search for dark
  matter}}, \href{https://doi.org/10.1038/s41586-018-0542-z}{\emph{Nature}
  {\bfseries 562} (2018) 51}
  [\href{https://arxiv.org/abs/1810.01668}{{\ttfamily 1810.01668}}].

\bibitem{xenon2018}
{\scshape XENON Collaboration 7} collaboration, \emph{Dark matter search
  results from a one ton-year exposure of xenon1t},
  \href{https://doi.org/10.1103/PhysRevLett.121.111302}{\emph{Physical Review
  Letters} {\bfseries 121} (2018) 111302}.

\bibitem{Schumann_2019}
M.~Schumann, \emph{Direct detection of wimp dark matter: concepts and status},
  \href{https://doi.org/10.1088/1361-6471/ab2ea5}{\emph{Journal of Physics G:
  Nuclear and Particle Physics} {\bfseries 46} (2019) 103003}.

\bibitem{Ackermann_2016}
{\scshape Fermi-LAT Collaboration} collaboration, \emph{Search for gamma-ray
  emission from the coma cluster with six years of fermi-lat data},
  \href{https://doi.org/10.3847/0004-637x/819/2/149}{\emph{The Astrophysical
  Journal} {\bfseries 819} (2016) 149}.

\bibitem{Albert_2017}
{\scshape Fermi-LAT and DES Collaborations} collaboration, \emph{Searching for
  dark matter annihilation in recently discovered milky way satellites with
  fermi-lat}, \href{https://doi.org/10.3847/1538-4357/834/2/110}{\emph{The
  Astrophysical Journal} {\bfseries 834} (2017) 110}.

\bibitem{Colafrancesco2006}
S.~Colafrancesco, S.~Profumo and P.~Ullio, \emph{Multi-frequency analysis of
  neutralino dark matter annihilations in the coma cluster}, {\emph{Astronomy
  and Astrophysics} {\bfseries 455} (2006) 21}.

\bibitem{Colafrancesco2007}
S.~Colafrancesco, S.~Profumo and P.~Ullio, \emph{Detecting dark matter wimps in
  the draco dwarf: a multi-wavelength perspective}, {\emph{Physical Review D}
  {\bfseries 75} (2007) 023513}.

\bibitem{beckm312019}
G.~Beck, \emph{{An excess of excesses examined via dark matter radio emissions
  from galaxies}},
  \href{https://doi.org/10.1088/1475-7516/2019/08/019}{\emph{JCAP} {\bfseries
  1908} (2019) 019} [\href{https://arxiv.org/abs/1905.05599}{{\ttfamily
  1905.05599}}].

\bibitem{regis2017}
M.~Regis, L.~Richter and S.~Colafrancesco, \emph{Dark matter in the reticulum
  ii dsph: a radio search}, {\emph{Journal of Cosmology and Astroparticle
  Physics} {\bfseries 2017} (2017) 025}.

\bibitem{Chan_2019}
M.~H. Chan, L.~Cui, J.~Liu and C.~S. Leung, \emph{"ruling out ~100-300 gev
  thermal relic annihilating dark matter by radio observation of the andromeda
  galaxy"}, \href{https://doi.org/10.3847/1538-4357/aafe0b}{\emph{The
  Astrophysical Journal} {\bfseries 872} (2019) 177}.

\bibitem{chan2017}
M.~H. Chan, \emph{{Constraining annihilating dark matter by radio data of
  M33}}, \href{https://doi.org/10.1103/PhysRevD.96.043009}{\emph{Physical
  Review} {\bfseries D96} (2017) 043009}
  [\href{https://arxiv.org/abs/1708.01370}{{\ttfamily 1708.01370}}].

\bibitem{HESS:2016ygi}
{\scshape H.E.S.S.} collaboration, \emph{{Search for dark matter annihilations
  towards the inner Galactic halo from 10 years of observations with H.E.S.S}},
  \href{https://doi.org/10.1103/PhysRevLett.117.111301}{\emph{Physical Review
  Letters} {\bfseries 117} (2016) 111301}
  [\href{https://arxiv.org/abs/1607.08142}{{\ttfamily 1607.08142}}].

\bibitem{egorov2013}
A.~E. {Egorov} and E.~{Pierpaoli}, \emph{{Constraints on dark matter
  annihilation by radio observations of M31}},
  \href{https://doi.org/10.1103/PhysRevD.88.023504}{\emph{Physical Review D}
  {\bfseries 88} (2013) 023504}
  [\href{https://arxiv.org/abs/1304.0517}{{\ttfamily 1304.0517}}].

\bibitem{FermiSMC2016}
R.~Caputo, M.~R. Buckley, P.~Martin, E.~Charles, A.~M. Brooks, A.~Drlica-Wagner
  et~al., \emph{Search for gamma-ray emission from dark matter annihilation in
  the small magellanic cloud with the fermi large area telescope},
  \href{https://doi.org/10.1103/PhysRevD.93.062004}{\emph{Physical Review D}
  {\bfseries 93} (2016) 062004}.

\bibitem{siffert2011}
B.~B. {Siffert}, A.~{Limone}, E.~{Borriello}, G.~{Longo} and G.~{Miele},
  \emph{{Radio emission from dark matter annihilation in the Large Magellanic
  Cloud}},
  \href{https://doi.org/10.1111/j.1365-2966.2010.17613.x}{\emph{Monthly Notices
  of the Royal Astronomical Society} {\bfseries 410} (2011) 2463}
  [\href{https://arxiv.org/abs/1006.5325}{{\ttfamily 1006.5325}}].

\bibitem{storm2017}
E.~Storm, T.~E. Jeltema, M.~Splettstoesser and S.~Profumo, \emph{Synchrotron
  emission from dark matter annihilation: Predictions for constraints from
  non-detections of galaxy clusters with new radio surveys}, {\emph{The
  Astrophysical Journal} {\bfseries 839} (2017) 33}.

\bibitem{gs2016}
G.~Beck and S.~Colafrancesco, \emph{{A Multi-frequency analysis of dark matter
  annihilation interpretations of recent anti-particle and $\gamma$-ray
  excesses in cosmic structures}},
  \href{https://doi.org/10.1088/1475-7516/2016/05/013}{\emph{Journal of
  Cosmology and Astroparticle Physics} {\bfseries 1605} (2016) 013}
  [\href{https://arxiv.org/abs/1508.01386}{{\ttfamily 1508.01386}}].

\bibitem{Beck_2019}
G.~Beck, \emph{An excess of excesses examined via dark matter radio emissions
  from galaxies},
  \href{https://doi.org/10.1088/1475-7516/2019/08/019}{\emph{Journal of
  Cosmology and Astroparticle Physics} {\bfseries 2019} (2019) 019}.

\bibitem{Boyarsky_2019}
A.~Boyarsky, M.~Drewes, T.~Lasserre, S.~Mertens and O.~Ruchayskiy,
  \emph{Sterile neutrino dark matter},
  \href{https://doi.org/10.1016/j.ppnp.2018.07.004}{\emph{Progress in Particle
  and Nuclear Physics} {\bfseries 104} (2019) 1}.

\bibitem{Adhikari_2017}
R.~Adhikari, M.~Agostini, N.~A. Ky, T.~Araki, M.~Archidiacono, M.~Bahr et~al.,
  \emph{A white paper on kev sterile neutrino dark matter},
  \href{https://doi.org/10.1088/1475-7516/2017/01/025}{\emph{Journal of
  Cosmology and Astroparticle Physics} {\bfseries 2017} (2017) 025}.

\bibitem{Duffy_2009}
L.~D. Duffy and K.~v. Bibber, \emph{Axions as dark matter particles},
  \href{https://doi.org/10.1088/1367-2630/11/10/105008}{\emph{New Journal of
  Physics} {\bfseries 11} (2009) 105008}.

\bibitem{axions2015}
P.~W. Graham, I.~G. Irastorza, S.~K. Lamoreaux, A.~Lindner and K.~A. van
  Bibber, \emph{Experimental searches for the axion and axion-like particles},
  \href{https://doi.org/10.1146/annurev-nucl-102014-022120}{\emph{Annual Review
  of Nuclear and Particle Science} {\bfseries 65} (2015) 485}
  [\href{https://arxiv.org/abs/https://doi.org/10.1146/annurev-nucl-102014-022120}{{\ttfamily
  https://doi.org/10.1146/annurev-nucl-102014-022120}}].

\bibitem{cast2017}
{\scshape CAST} collaboration, \emph{New cast limit on the axion-photon
  interaction}, \href{https://doi.org/10.1038/nphys4109}{\emph{Nature Physics}
  {\bfseries 13} (2017) 584}.

\bibitem{Braine_2020}
T.~Braine, R.~Cervantes, N.~Crisosto, N.~Du, S.~Kimes, L.~Rosenberg et~al.,
  \emph{Extended search for the invisible axion with the axion dark matter
  experiment},
  \href{https://doi.org/10.1103/physrevlett.124.101303}{\emph{Physical Review
  Letters} {\bfseries 124} (2020) }.

\bibitem{Bashkanov_2020}
M.~Bashkanov and D.~P. Watts, \emph{A new possibility for light-quark dark
  matter}, \href{https://doi.org/10.1088/1361-6471/ab67e8}{\emph{Journal of
  Physics G: Nuclear and Particle Physics} {\bfseries 47} (2020) 03LT01}.

\bibitem{Ackermann_2017gc}
M.~Ackermann, M.~Ajello, A.~Albert, W.~B. Atwood, L.~Baldini, J.~Ballet et~al.,
  \emph{The fermi galactic center gev excess and implications for dark matter},
  \href{https://doi.org/10.3847/1538-4357/aa6cab}{\emph{The Astrophysical
  Journal} {\bfseries 840} (2017) 43}.

\bibitem{bonnivard2015}
V.~Bonnivard, C.~Combet, D.~Maurin, A.~Geringer-Sameth, S.~M. Koushiappas,
  M.~G. Walker et~al., \emph{{Dark matter annihilation and decay profiles for
  the Reticulum II dwarf spheroidal galaxy}},
  \href{https://doi.org/10.1088/2041-8205/808/2/L36}{\emph{The Astrophysical
  Journal} {\bfseries 808} (2015) L36}
  [\href{https://arxiv.org/abs/1504.03309}{{\ttfamily 1504.03309}}].

\bibitem{beckm31}
R.~Beck, \emph{The magnetic field in m31}, {\emph{Astronomy and Astrophysics}
  {\bfseries 106} (1982) 121}.

\bibitem{briel1992}
U.~G. {Briel}, J.~P. {Henry} and H.~{Boehringer}, \emph{{Observation of the
  Coma cluster of galaxies with ROSAT during the all-sky-survey.}},
  {\emph{Astronomy and Astrophysics} {\bfseries 259} (1992) L31}.

\bibitem{bonafede2010}
A.~Bonafede, L.~Feretti, M.~Murgia, F.~Govoni, G.~Giovannini, D.~Dallacasa
  et~al., \emph{The coma cluster magnetic field from faraday rotation
  measures}, \href{https://doi.org/10.1051/0004-6361/200913696}{\emph{Astronomy
  and Astrophysics} {\bfseries 513} (2010) A30}.

\bibitem{ruiz-granados2010}
B.~Ruiz-Granados, J.~A. Rubi{\~{n}}o-Mart{\'{\i}}n, E.~Florido and E.~Battaner,
  \emph{Magnetic fields and the outer rotation curve of m31},
  \href{https://doi.org/10.1088/2041-8205/723/1/l44}{\emph{The Astrophysical
  Journal} {\bfseries 723} (2010) L44}.

\bibitem{tamm2012}
A.~{Tamm}, E.~{Tempel}, P.~{Tenjes}, O.~{Tihhonova} and T.~{Tuvikene},
  \emph{{Stellar mass map and dark matter distribution in M 31}},
  \href{https://doi.org/10.1051/0004-6361/201220065}{\emph{Astronomy and
  Astrophysics} {\bfseries 546} (2012) A4}
  [\href{https://arxiv.org/abs/1208.5712}{{\ttfamily 1208.5712}}].

\bibitem{clumpy_2012}
A.~Charbonnier, C.~Combet and D.~Maurin, \emph{Clumpy: A code for
  {\ensuremath{\gamma}}-ray signals from dark matter structures},
  \href{https://doi.org/10.1016/j.cpc.2011.10.017}{\emph{Computer Physics
  Communications} {\bfseries 183} (2012) 656}.

\bibitem{clumpy_2016}
V.~Bonnivard, M.~H{\"u}tten, E.~Nezri, A.~Charbonnier, C.~Combet and D.~Maurin,
  \emph{{CLUMPY : Jeans analysis, {\ensuremath{\gamma}}-ray and
  {\ensuremath{\nu}} fluxes from dark matter (sub-)structures}},
  \href{https://doi.org/10.1016/j.cpc.2015.11.012}{\emph{Computer Physics
  Communications} {\bfseries 200} (2016) 336}.

\bibitem{clumpy_2018}
M.~{H{\"u}tten}, C.~{Combet} and D.~{Maurin}, \emph{{CLUMPY v3:
  {\ensuremath{\gamma}}-ray and {\ensuremath{\nu}} signals from dark matter at
  all scales}}, \href{https://doi.org/10.1016/j.cpc.2018.10.001}{\emph{Computer
  Physics Communications} {\bfseries 235} (2019) 336}
  [\href{https://arxiv.org/abs/1806.08639}{{\ttfamily 1806.08639}}].

\bibitem{Essig:2013goa}
R.~Essig, E.~Kuflik, S.~D. McDermott, T.~Volansky and K.~M. Zurek,
  \emph{{Constraining Light Dark Matter with Diffuse X-Ray and Gamma-Ray
  Observations}}, \href{https://doi.org/10.1007/JHEP11(2013)193}{\emph{JHEP}
  {\bfseries 11} (2013) 193} [\href{https://arxiv.org/abs/1309.4091}{{\ttfamily
  1309.4091}}].

\bibitem{Liu_2016}
H.~Liu, T.~R. Slatyer and J.~Zavala, \emph{Contributions to cosmic reionization
  from dark matter annihilation and decay},
  \href{https://doi.org/10.1103/physrevd.94.063507}{\emph{Physical Review D}
  {\bfseries 94} (2016) }.

\bibitem{scanlon1965}
J.~H. {Scanlon} and S.~N. {Milford}, \emph{{Energy Spectra of Electrons from
  {\ensuremath{\pi}}-{\ensuremath{\mu}}-e Decays in Interstellar Space.}},
  \href{https://doi.org/10.1086/148156}{\emph{The Astrophysical Journal}
  {\bfseries 141} (1965) 718}.

\bibitem{Gal:2019hdb}
A.~Gal, \emph{{Pion assisted dibaryons: the d*(2380) resonance}},
  \href{https://doi.org/10.1051/epjconf/201919902018}{\emph{EPJ Web Conf.}
  {\bfseries 199} (2019) 02018}.

\bibitem{baltz1999}
E.~A. Baltz and J.~Edsj\"o, \emph{Positron propagation and fluxes from
  neutralino annihilation in the halo},
  \href{https://doi.org/10.1103/PhysRevD.59.023511}{\emph{Physical Review D}
  {\bfseries 59} (1998) 023511}.

\bibitem{baltz2004}
E.~A. Baltz and L.~Wai, \emph{Diffuse inverse compton and synchrotron emission
  from dark matter annihilations in galactic satellites},
  \href{https://doi.org/10.1103/PhysRevD.70.023512}{\emph{Physical Review D}
  {\bfseries 70} (2004) 023512}.

\bibitem{longair1994}
M.~S. Longair, \emph{High Energy Astrophysics}. Cambridge University Press,
  1994.

\bibitem{rybicki1986}
G.~B. {Rybicki} and A.~P. {Lightman}, \emph{{Radiative Processes in
  Astrophysics}}. Wiley, June, 1986.

\bibitem{nfw1996}
J.~F. Navarro, C.~S. Frenk and S.~D.~M. White, \emph{{The Structure of cold
  dark matter halos}}, \href{https://doi.org/10.1086/177173}{\emph{The
  Astrophysical Journal} {\bfseries 462} (1996) 563}
  [\href{https://arxiv.org/abs/astro-ph/9508025}{{\ttfamily
  astro-ph/9508025}}].

\bibitem{burkert1995}
A.~Burkert, \emph{{The Structure of dark matter halos in dwarf galaxies}},
  \href{https://doi.org/10.1086/309560}{\emph{IAU Symp.} {\bfseries 171} (1996)
  175} [\href{https://arxiv.org/abs/astro-ph/9504041}{{\ttfamily
  astro-ph/9504041}}].

\bibitem{einasto1968}
J.~Einasto, \emph{On galactic descriptive functions}, {\emph{Publications of
  the Tartuskoj Astrofizica Observatory} {\bfseries 36} (1968) 414}.

\bibitem{coma-halo-2003}
E.~L. \L{}okas and G.~A. Mamon, \emph{{Dark matter distribution in the Coma
  cluster from galaxy kinematics: breaking the mass-anisotropy degeneracy}},
  \href{https://doi.org/10.1046/j.1365-8711.2003.06684.x}{\emph{Monthly Notices
  of the Royal Astronomical Society} {\bfseries 343} (2003) 401}.

\bibitem{chen-clusters-2007}
Y.~{Chen}, T.~H. {Reiprich}, H.~{B{\"o}hringer}, Y.~{Ikebe} and Y.~Y. {Zhang},
  \emph{{Statistics of X-ray observables for the cooling-core and non-cooling
  core galaxy clusters}},
  \href{https://doi.org/10.1051/0004-6361:20066471}{\emph{Astronomy and
  Astrophysics} {\bfseries 466} (2007) 805}
  [\href{https://arxiv.org/abs/astro-ph/0702482}{{\ttfamily
  astro-ph/0702482}}].

\bibitem{coma-radio2003}
M.~Thierbach, U.~Klein and R.~Wielebinski, \emph{The diffuse radio emission
  from the coma cluster at 2.675 ghz and 4.85 ghz}, {\emph{Astronomy and
  Astrophysics} {\bfseries 397} (2003) 53}.

\bibitem{Ackermann_2017}
{\scshape Fermi-LAT Collaboration} collaboration, \emph{Observations of m31 and
  m33 with the fermi large area telescope: A galactic center excess in
  andromeda?}, \href{https://doi.org/10.3847/1538-4357/aa5c3d}{\emph{The
  Astrophysical Journal} {\bfseries 836} (2017) 208}.

\bibitem{walker2009}
M.~G. Walker, M.~Mateo, E.~W. Olszewski, J.~P. narrubia, N.~W. Evans and
  G.~Gilmore, \emph{A universal mass profile for dwarf spheroidal galaxies?},
  {\emph{The Astrophysical Journal} {\bfseries 704} (2009) 1274}.

\bibitem{adams2014}
J.~J. Adams et~al., \emph{Dwarf galaxy dark matter density profiles inferred
  from stellar and gas kinematics}, {\emph{The Astrophysical Journal}
  {\bfseries 789} (2014) 63}.

\bibitem{bechtol2015}
{\scshape DES Collaboration} collaboration, \emph{{Eight New Milky Way
  Companions Discovered in First-year Dark Energy Survey Data}},
  \href{https://doi.org/10.1088/0004-637X/807/1/50}{\emph{The Astrophysical
  Journal} {\bfseries 807} (2015) 50}
  [\href{https://arxiv.org/abs/1503.02584}{{\ttfamily 1503.02584}}].

\bibitem{koposov2015}
S.~E. Koposov, V.~Belokurov, G.~Torrealba and N.~W. Evans, \emph{{Beasts of the
  Southern Wild: Discovery of nine Ultra Faint satellites in the vicinity of
  the Magellanic Clouds}},
  \href{https://doi.org/10.1088/0004-637X/805/2/130}{\emph{The Astrophysical
  Journal} {\bfseries 805} (2015) 130}
  [\href{https://arxiv.org/abs/1503.02079}{{\ttfamily 1503.02079}}].

\bibitem{g4002013}
A.~M. Galper, O.~Adriani, R.~L. Aptekar, I.~V. Arkhangelskaja, A.~I.
  Arkhangelskiy, M.~Boezio et~al., \emph{Design and performance of the
  gamma-400 gamma-ray telescope for dark matter searches}, .

\bibitem{g4002019}
N.~P. {Topchiev}, A.~M. {Galper}, I.~V. {Arkhangelskaja}, A.~I.
  {Arkhangelskiy}, A.~V. {Bakaldin}, I.~V. {Chernysheva} et~al.,
  \emph{{High-energy gamma- and cosmic-ray observations with future space-based
  GAMMA-400 gamma-ray telescope}},  in \emph{European Physical Journal Web of
  Conferences}, vol.~208 of \emph{European Physical Journal Web of
  Conferences}, p.~14004, May, 2019,
  \href{https://doi.org/10.1051/epjconf/201920814004}{DOI}.

\bibitem{g4002020}
A.~E. Egorov, N.~P. Topchiev, A.~M. Galper, O.~D. Dalkarov, A.~A. Leonov, S.~I.
  Suchkov et~al., \emph{Dark matter searches by the planned gamma-ray telescope
  gamma-400},  2020.

\bibitem{Amati_2018}
L.~Amati, P.~O'Brien, D.~G\"{o}tz, E.~Bozzo, C.~Tenzer, F.~Frontera et~al.,
  \emph{The theseus space mission concept: science case, design and expected
  performances},
  \href{https://doi.org/10.1016/j.asr.2018.03.010}{\emph{Advances in Space
  Research} {\bfseries 62} (2018) 191}.

\bibitem{Chan_2020}
M.~H. Chan, \emph{The decaying and scattering properties of the d*(2380)
  hexaquark bose-einstein condensate dark matter},
  \href{https://doi.org/10.3847/1538-4357/ab9df6}{\emph{The Astrophysical
  Journal} {\bfseries 898} (2020) 132}.

\end{thebibliography}\endgroup

\end{document}